# Magnetic structures of rare earth intermetallic compounds RCuAs$_2$ (R = Pr, Nd, Tb, Dy, Ho, and Yb)

Y Zhao [a, b, *], J W Lynn [a], G S Thakur [c,1], Z Haque [c], L C Gupta [c,2] and A K Ganguli [c,d]

[a] NIST Center for Neutron Research, National Institute of Standards and Technology, Gaithersburg, MD, 20899, USA

[b] Department of Materials Science and Engineering, University of Maryland, College Park, MD, 20742, USA

[c] Department of Chemistry, Indian Institute of Technology Delhi, 110016, India

[d] Institute of Nano Science and Technology, Mohali, 160064, India

**Abstract**

Neutron scattering studies have been carried out on polycrystalline samples of a series of rare earth intermetallic compounds RCuAs$_2$ (R = Pr, Nd, Dy, Tb, Ho and Yb) as a function of temperature to determine the magnetic structures and the order parameters. These compounds crystallize in the ZrCuSi$_2$ type structure, which is similar to that of the RFeAsO (space group $P4/nmm$) class of iron-based superconductors. PrCuAs$_2$ develops commensurate magnetic order with $\boldsymbol{K}$ = (0, 0, 0.5) below T$_N$ = 6.4(2) K, with the ordered moments pointing along the $\boldsymbol{c}$-axis. The irreducible representation analysis shows either a $\Gamma^1_2$ or $\Gamma^1_3$ representation. NdCuAs$_2$ and DyCuAs$_2$ order below T$_N$ = 3.54(5) K and T$_N$ = 10.1(2) K, respectively, with the same ordering wave vector but the moments lying in the $\boldsymbol{a}$-$\boldsymbol{b}$ plane (with a $\Gamma^2_9$ or $\Gamma^2_{10}$ representation). TbCuAs$_2$ and HoCuAs$_2$ exhibit incommensurate magnetic structures below T$_N$ = 9.44(7) and 4.41(2) K, respectively. For TbCuAs$_2$, two separate magnetic ordering wave vectors are established as $\boldsymbol{K}_{1(Tb)}$ = (0.240, 0.155, 0.48) and $\boldsymbol{K}_{2(Tb)}$ = (0.205, 0.115, 0.28), whereas HoCuAs$_2$ forms a single $\boldsymbol{K}_{(Ho)}$ = (0.121, 0.041, 0.376) magnetic structure with 3$^{rd}$ order harmonic magnetic peaks. YbCuAs$_2$ does not exhibit any magnetic Bragg peaks at 1.5 K, while susceptibility measurements indicate an antiferromagnetic-like transition at 4 K, suggesting that either the ordering is not long range in nature or the ordered moment is below the sensitivity limit of ≈ 0.2 $\mu_B$.

---

\* Corresponding author
  *Email address:* yang.zhao@nist.gov (Y. Zhao).
[1] Current address: Max Planck Institute for Chemical Physics of Solids, Nöthnitzer Strasse 40, 01187, Dresden, Germany
[2] Visiting scientist at department of chemistry IIT Delhi

## 1. Introduction

Rare earth (R) compounds have generated considerable interest over many decades because of the rich variety of exotic phenomena they display [1], such as unconventional superconductivity[2], coexistence of superconductivity and magnetism [3], multiferroicity [4–6], heavy fermion behaviour[7], and high-$T_c$ superconductivity in cuprates [8]. The most recent exotic phenomenon in rare earth compounds is the high-temperature superconductivity in iron pnictides such as $RFeAsO_{1-x}F_x$ and related systems[9]. The present $RCuAs_2$ compounds are a new class of rare earth intermetallic materials, which have been the subject of recent studies that have revealed several anomalous properties [10–12]. In these materials, the magnetic properties arise only from the $R^{3+}$ ions, as the $Cu^{1+}$ ions are non-magnetic ($d^{10}$ system). Interestingly, several compounds of $RCuAs_2$ (R = Tb, Dy, Sm, and Gd) show an anomalous resistivity minimum which happens well above the magnetic phase transition to long range magnetic order, while others (R = Pr, Nd, Ho, and Yb) do not have such a resistivity minimum even though the magnetic phase transitions are still present. Although the origin of this anomalous resistivity minima is not fully understood, it is believed to be related to the magnetism in these materials. Here we report comprehensive neutron scattering studies of the crystallographic and magnetic properties of the polycrystalline $RCuAs_2$ (R = Pr, Nd, Dy, Tb, Ho, and Yb) series of compounds. $CeCuAs_2$ was not considered in these studies as this material shows no evidence of any magnetic order for temperatures down to 45 mK as reported in an earlier work. [13] We confirm that all the measured compounds share the same crystal structure, while the magnetic structures are varied, including both commensurate and incommensurate (IC) antiferromagnetic orderings. Based on the observed magnetic diffraction patterns, we discuss the possible magnetic structures, order parameters, and ordered magnetic moments for this new class of intermetallic compounds.

## 2. Experiment Details

Polycrystalline $RCuAs_2$ samples were prepared by employing the solid state synthesis technique. Powders of pure Cu and As and lumps of rare earth metal (purity > 99.5 %) were weighed and mixed thoroughly in an argon filled glove box (with $H_2O$, $O_2$ < 0.1 μg/g). The mixture was then sealed in evacuated quartz tubes under dynamic vacuum. The ampoules were heated at 900 °C for a week. After an intermediate grinding, additional sintering under the same conditions was performed to obtain homogenous good quality samples. The phase purity of all the samples was checked by powder x-ray diffraction technique using Cu-$K_α$ radiation in the 2θ range of 10° to 70° at room temperature. Magnetization measurements in the temperature range from 2 K to 300 K with applied external magnetic field of $μ_0H$ = 0.5 T were performed on all the samples using a Superconducting Quantum Interference Device (SQUID) magnetometer.

Neutron diffraction experiments were carried out by using the BT1 high-resolution powder diffractometer and BT7 thermal triple-axis spectrometer at National Institute of Standard and Technology (NIST) Center for Neutron Research. All the polycrystalline samples of RCuAs$_2$ (R = Pr, Nd, Dy, Tb, Ho, and Yb) used for these neutron scattering measurements weighed around 1 gram. The samples were sealed in either Aluminium (BT7) or Vanadium (BT1) cans and cooled to the base temperature, which is around 0.4 K to 2.5 K depending upon the type of the cryogenics required. For measurements on BT1, the incident neutron wavelength was $\lambda = 1.5397$ Å using the Cu (311) monochromator and collimations of 60′ - 15′ - 7′ full-width-at-half-maximum (FWHM) before and after the monochromator and after the sample, respectively. For coarse-resolution/high-intensity diffraction measurements on BT7 the incident neutron wavelength employed was $\lambda = 2.359$ Å and the position sensitive detector (PSD) was placed in the straight-through position to cover around 5 degrees in scattering $2\theta$ angles, taking data in steps of 0.25° and then binning the data to obtain the diffraction pattern [14]. PG filters were employed along the neutron beam path before the monochromator and after the sample to eliminate higher order wavelength contaminations. Unless specified, the collimations used in BT7 measurements were open - 80′ - 80′ radial (FWHM) along the neutron beam path before and after the monochromator, and after sample, respectively. All the neutron powder diffraction data have been refined using the Rietveld method with software suite *FullProf*[15]. Representation analysis was carried out using the software *SARAh* [16] to determine the symmetry-consistent irreducible representations (IRs) of the magnetic structures.

3. Results and Discussion

3.1 Crystal Structure

All the RCuAs$_2$ compounds share the same crystal structure (ZrCuSi$_2$ type with space group *P*4/*nmm*) shown in figure 1, confirmed by Rietveld refinement results from neutron powder diffraction measurements. We did not observe any indication of structural phase transitions in any of the compounds in the temperature range from room temperature to the lowest measured temperature. Detailed refinement results are listed in Table I. The crystal structure contains a sequence of layers by means of As$_{(1)}$ – R - As$_{(2)}$ – Cu - As$_{(2)}$ – R – As$_{(1)}$, where As$_{(1)}$ and As$_{(2)}$ are the two inequivalent As-sites, with distorted edge-sharing Cu-As$_{(2)}$ tetrahedral alternating with distorted As$_{(1)}$-R-As$_{(2)}$ square antiprisms. The sites refined as fully occupied within uncertainties, and in the final fits, the occupancies were fixed at the fully occupied values. These results are consistent with the previously reported x-ray powder diffraction measurements on these materials [17–19]. We did notice that small amounts of impurities (< 5 %), which are in the form of *R*As and/or Cu$_3$As/Cu$_5$As$_2$, exist and were detected by neutron diffraction. However, for the crystal structure refinements, most of the impurity peaks are well separated from the majority phase of RCuAs$_2$ Bragg peaks and can be excluded from the refinement process. In addition, when an Al sample holder was employed there were Al powder peaks. These peaks are well known and were co-refined with the RCuAs$_2$, and the quoted uncertainties include both effects.

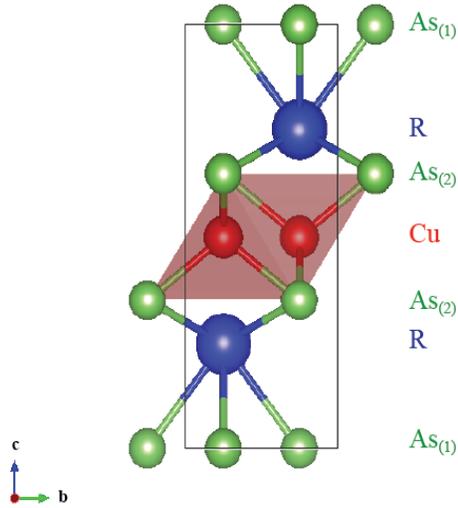

Figure 1. (colour online) Crystal structure of RCuAs$_2$ depicting the edge shared Cu – As$_{(2)}$ tetrahedron and distorted As$_{(1)}$ – R – As$_{(2)}$ square antiprism. The solid (black) box marks the structural unit cell. The figure was created using the VESTA program[22].

Table I. Summary of the structure refinement results for RCuAs$_2$ at temperatures above the magnetic phase transition, T$_N$. All the compounds share the same P4/nmm (# 129) space group, and we did not observe any indication of structural distortions associated with the magnetic transition.

| Compounds | PrCuAs$_2$ | NdCuAs$_2$ | TbCuAs$_2$ | DyCuAs$_2$ | HoCuAs$_2$ | YbCuAs$_2$ |
|---|---|---|---|---|---|---|
| Instrument | BT7 | BT7 | BT1 | BT7 | BT1 | BT7 |
| λ (Å) | 2.359 | 2.359 | 1.5397 | 2.359 | 1.5397 | 2.359 |
| T (K) | 10 | 10 | 11 | 20 | 4.5 | 20 |
| a(Å) | 3.9806 (3) | 3.9534 (4) | 3.8818 (1) | 3.908 (6) | 3.8609 (1) | 3.8414 (9) |
| c(Å) | 10.065 (1) | 10.029 (3) | 9.8671 (4) | 9.89 (3) | 9.7997 (4) | 9.729 (9) |
| R(z) [a] | 0.246 (1) | 0.241 (1) | 0.2394 (5) | 0.236 (7) | 0.2368 (9) | 0.231 (1) |
| As2(z) | 0.3502 (6) | 0.346 (1) | 0.3381 (5) | 0.38 (1) | 0.333 (1) | 0.332 (2) |

[a] With R(z) and As2(z) as given above, the full set of the fractional coordinates of all the atoms are: R (R = Pr, Nd, Tb, Dy, Ho, and Yb): (0.25, 0.25, R(z)); As1: (0.75, 0.25, 0.0); As2: (0.75, 0.75, As2(z)); and Cu: (0.75, 0.25, 0.5).

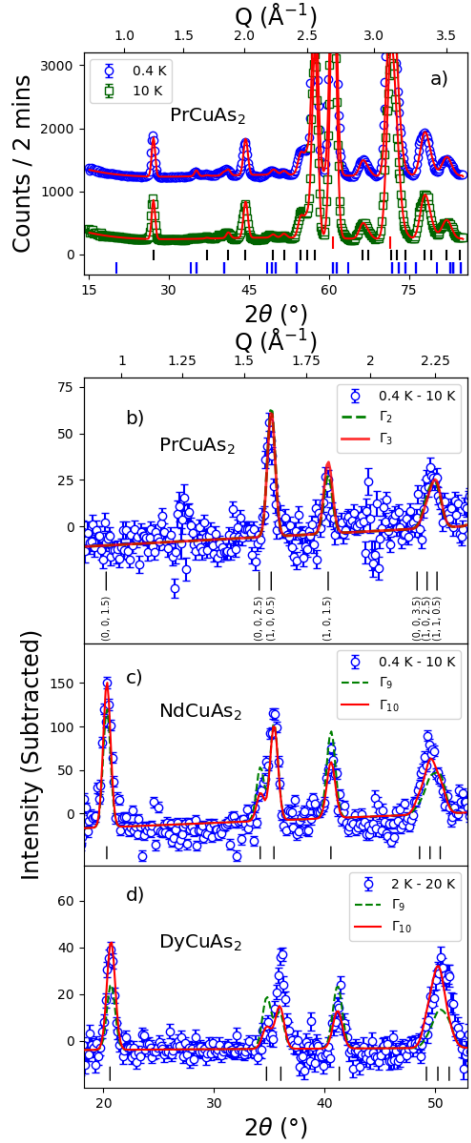

Figure 2 (colour online) (a) Neuron powder diffraction patterns of $PrCuAs_2$ at T = 0.4 K (blue circles, vertically offset by 1000) and 10 K (green squares) collected on the BT7 instrument. The solid (red) lines are the Rietveld refinement results. The vertical bars are the Bragg Peak positions for (from top to bottom) Al {111} and {200} from sample holder (red), $PrCuAs_2$ structural (black), and magnetic (blue). (b) – (d) Magnetic (subtracted) diffraction patterns for $PrCuAs_2$, $NdCuAs_2$, and $DyCuAs_2$, respectively. The dashed (green) and solid (red) lines are magnetic diffraction pattern calculations based on two different magnetic models. The vertical bars are the magnetic Bragg peak positions for each of the compounds. The data are plotted against $2\theta$ (bottom) and corresponding Q values (top). Error bars in all figures are statistical in origin and represent one standard deviation.

### 3.2 Commensurate Magnetic Order in $PrCuAs_2$, $NdCuAs_2$, and $DyCuAs_2$

We start our discussion with the data for $PrCuAs_2$, $NdCuAs_2$, and $DyCuAs_2$, which exhibit simple commensurate magnetic structures. Figure 2(a) shows representative neutron diffraction data measured on BT7 and Rietveld refinement results for the powder $PrCuAs_2$ sample collected at 0.4 K and 10 K. For unpolarized neutrons the nuclear

Table II. Basis vectors (BVs) for the space group P4/nmm (#129) with K = (0, 0, 0.5). The decomposition of the magnetic representation for the R (R = Pr, Nd, and Dy) site, including only non-zero IRs, is $\Gamma_{Mag} = \Gamma^1_2 + \Gamma^1_3 + \Gamma^2_9 + \Gamma^2_{10}$. The atoms of the nonprimitive basis are defined according to 1: (0.25, 0.25, 0.246), 2: (0.75, 0.75, 0.754).

| IR | BV | Atom | BV components | | | | | |
|---|---|---|---|---|---|---|---|---|
| | | | $m_a$ | $m_b$ | $m_c$ | $im_a$ | $im_b$ | $im_c$ |
| $\Gamma_2$ | $\psi_1$ | 1 | 0 | 0 | 8 | 0 | 0 | 0 |
| | | 2 | 0 | 0 | 8 | 0 | 0 | 0 |
| $\Gamma_3$ | $\psi_2$ | 1 | 0 | 0 | 8 | 0 | 0 | 0 |
| | | 2 | 0 | 0 | -8 | 0 | 0 | 0 |
| $\Gamma_9$ | $\psi_3$ | 1 | 4 | 0 | 0 | 0 | 0 | 0 |
| | | 2 | -4 | 0 | 0 | 0 | 0 | 0 |
| | $\psi_4$ | 1 | 0 | -4 | 0 | 0 | 0 | 0 |
| | | 2 | 0 | 4 | 0 | 0 | 0 | 0 |
| $\Gamma_{10}$ | $\psi_5$ | 1 | 0 | 4 | 0 | 0 | 0 | 0 |
| | | 2 | 0 | 4 | 0 | 0 | 0 | 0 |
| | $\psi_6$ | 1 | 4 | 0 | 0 | 0 | 0 | 0 |
| | | 2 | 4 | 0 | 0 | 0 | 0 | 0 |

and magnetic Bragg intensities simply add, and hence the nuclear Bragg peaks are eliminated in the subtraction (assuming a negligible structural distortion associated with the ordering), revealing the magnetic diffraction pattern. Figure 2(b) – (d) shows the subtracted intensities for PrCuAs$_2$, NdCuAs$_2$, and DyCuAs$_2$, which were collected below and well above the antiferromagnetic ordering temperature for each of these compounds. In the angular range (18 ° < $2\theta$ < 53°) shown in figure 2b)-d), we find magnetic peaks that can be indexed as the Q$_{mag}$ = Q$_{str}$ + (0, 0, 0.5) in reciprocal lattice units (r. l. u.), where Q$_{mag}$ and Q$_{str}$ are Bragg peak positions for the magnetic and crystal structures, respectively. This establishes the magnetic propagation wave vector *K* as (0, 0, 0.5) for all three compounds. In addition, the systematic absence of the magnetic Bragg peaks at the Q = (0, 0, n + 0.5) positions for PrCuAs$_2$ (figure 2(b), where n = 0, 1, 2...), indicates that the magnetic moment direction of the Pr$^{3+}$ is along the *c*-axis, as only the magnetic components perpendicular to the momentum transfer Q can contribute to the magnetic Bragg diffraction intensities. The NdCuAs$_2$ and DyCuAs$_2$ data, on the other hand, show strong magnetic peaks at (0, 0, 1.5) (figure 2(c) and (d)), indicating that the ordered moment direction is in the *a-b* plane. We remark that it is not possible to determine the spin direction within the plane from powder neutron diffraction due to the tetragonal crystal

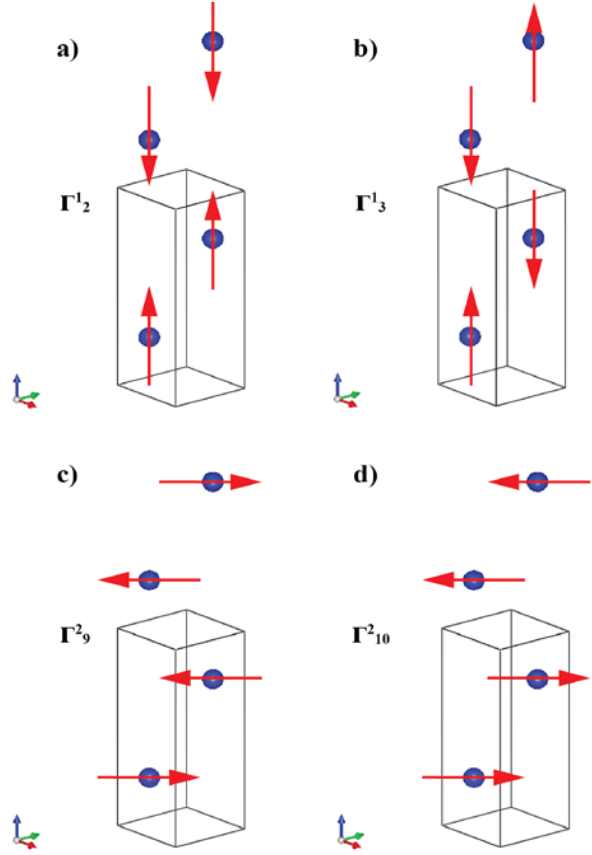

Figure 3 (colour online) Illustration of Irreducible representations of RCuAs$_2$ (only R ions are displayed) with K = (0, 0, 0.5) as (a) $\Gamma^1_2$, (b) $\Gamma^1_3$, (c) $\Gamma^2_9$, and (d) $\Gamma^2_{10}$.

symmetry. The observed difference in the magnetic moment orientations in these three materials is likely due to the crystal electrical field (CEF) effects.

To solve the magnetic structure, magnetic representation analyses were carried out with the software *Sarah*[16]. The decomposition of the IRs for the R$^{3+}$ ions with *K* = (0, 0, 0.5) can be written as $\Gamma_{Mag} = \Gamma^1_2 + \Gamma^1_3 + \Gamma^2_9 + \Gamma^2_{10}$. The basis vectors are listed in Table II and the illustration for the magnetic structures for each of the IRs is shown in figure 3. All four IRs can be considered as bilayer magnetic structures, where within the *a-b* plane the spins are aligned ferromagnetically, while the inter-plane interactions are alternating between ferromagnetic and antiferromagnetic depending on going through the R-As$_{(1)}$-R layers or R-As$_{(2)}$-Cu-As$_{(2)}$-R layers. For $\Gamma^1_2$ and $\Gamma^2_{10}$, the antiferromagnetic coupling happens between the R-As$_{(1)}$-R layers, while R – As$_{(2)}$ – Cu –As$_{(2)}$- R shows ferromagnetic coupling. In contrast, $\Gamma^1_3$ and $\Gamma^2_9$ demonstrate exactly the opposite behaviour. In addition, for $\Gamma^1_2$ (figure 3(a)) and $\Gamma^1_3$ (figure 3(b)), the ordered magnetic moment directions are along the c-axis, while for $\Gamma^2_9$ (figure 3(c)) and $\Gamma^2_{10}$ (figure 3(d)), the ordered moments lie in the *a-b* plane. Since we cannot determine the spin direction within *a-b* plane solely based on the neutron powder diffraction results, for convenience the spin directions were fixed along [110] during the magnetic structure refinement for the $\Gamma^2_9$ and $\Gamma^2_{10}$ IRs.

Table III. The refinement results of the ordered magnetic moments based on different IRs for PrCuAs$_2$, NdCuAs$_2$, and DyCuAs$_2$ at the lowest measured temperature.

|  | $\Gamma^1_2$ ($\mu_B$) | $\Gamma^1_3$ ($\mu_B$) | $\Gamma^2_9$ ($\mu_B$) | $\Gamma^2_{10}$ ($\mu_B$) |
| --- | --- | --- | --- | --- |
| **Pr (0.4 K)** | 1.15(6) | 1.17(6) | N/A | N/A |
| **Nd (0.4 K)** | N/A | N/A | 3.2(1) | 3.0(1) |
| **Dy (2.5 K)** | N/A | N/A | 10(1) | 10(1) |

Following the above discussion, since the magnetic moments of the Pr ions point along the *c*-axis, only the $\Gamma^1_2$ and $\Gamma^1_3$ representations are possible (figure 3(a) and (b)). For the Nd and Dy compounds the spins lie in *a-b* plane, so only $\Gamma^2_9$ and $\Gamma^2_{10}$ can be the possible options (figure 3(c) and 3(d)). Due to the powder average, $\Gamma^1_2$ and $\Gamma^1_3$ lead to virtually identical combined diffraction intensities. As a result, it is extremely difficult to distinguish the differences of the calculated diffraction pattern in carrying out the Rietveld refinement process. Even though the $\Gamma^1_3$ representation was indicative of slightly better fit for PrCuAs$_2$, and $\Gamma^2_{10}$ representation was indicative of slightly better fit for NdCuAs$_2$, we cannot completely rule out the possibility of the other magnetic structure purely based on the neutron powder diffraction results. The ordered magnetic moments given by the refinements are listed in Table III for each possibility. The ordered magnetic moment of Pr$^{3+}$ at 0.4 K is either 1.15(6) $\mu_B$ or 1.17(6) $\mu_B$ for $\Gamma^1_2$ or $\Gamma^1_3$, respectively, which are identical within the uncertainties. Note that the value of the ordered moment is much reduced from the free-ion Pr-ion ordered magnetic moment of 3.2 $\mu_B$ (=gJ) [1], indicating that there are significant CEF effects that can change the magnetic ground-state from its free-ion 4f-electron configuration. The Nd$^{3+}$ ordered magnetic moments at 0.4 K are 3.2(1) $\mu_B$ and 3.0(1) $\mu_B$ for $\Gamma^2_9$ and $\Gamma^2_{10}$, respectively, again the same within the uncertainties. Even though the Nd-ion ordered moments are close to the free-ion values for Nd$^{3+}$ of 3.27 $\mu_B$, the CEF still can have similar magnitude of effects. For DyCuAs$_2$, both the structure and magnetic Bragg peak intensities are much weaker due to the high absorption cross-section from Dy. Therefore, we can only estimate that the ordered Dy$^{3+}$ magnetic moment at 2.5 K is 10(1) $\mu_B$ for either $\Gamma^2_9$ or $\Gamma^2_{10}$, virtually identical to the free-ion value (10.0 $\mu_B$).[1] Further information on the magnetic structure and unambiguous distinction between the different magnetic models will require single crystal neutron diffraction measurements if high quality single crystal samples become available in the future.

Figure 4 shows measurements of the integrated magnetic Bragg peak intensities, proportional to the square of the antiferromagnetic order parameter (sublattice magnetization), at (1, 0, 0.5) for PrCuAs$_2$ and (0, 0, 1.5) for NdCuAs$_2$ and DyCuAs$_2$, respectively. The temperature-independent data collected well above the Néel temperatures (T$_N$) were used as background and subtracted from the data. The integrated intensities for each material smoothly emerge as the temperature drops below the T$_N$, indicating that the magnetic transition is 2$^{nd}$ order

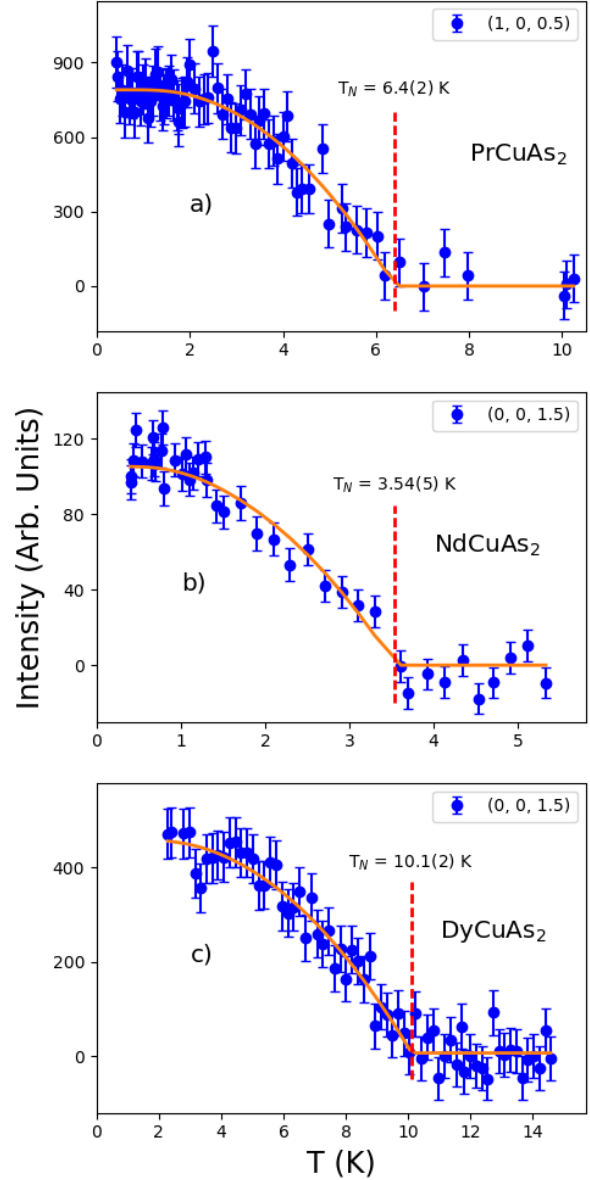

Figure 4 (colour online) Temperature dependence of the integrated magnetic Bragg peak intensities of (a) $PrCuAs_2$, (b) $NdCuAs_2$, and (c) $DyCuAs_2$. The solid (orange) curves show the mean field order parameter fit results for each of these compounds. The vertical dashed (red) lines indicate the fitted $T_N$ = 6.4(2) K for $PrCuAs_2$, $T_N$ = 3.54(5) K for the $NdCuAs_2$, and $T_N$ = 10.1(2) K for the $DyCuAs_2$.

in nature. The solid curves shown in figure 4 (a) – (c) are simple fits using mean field theory to fit the intensities, which are proportional to the square of the ordered magnetic moment. The fits give $T_N$ = 6.4(2) K for $PrCuAs_2$, $T_N$ = 3.54(5) K for $NdCuAs_2$, and $T_N$ = 10.1(2) K for $DyCuAs_2$. The above results are in very good agreement with the bulk magnetic property measurements reported previously[10,11] and in this work as well (not shown).

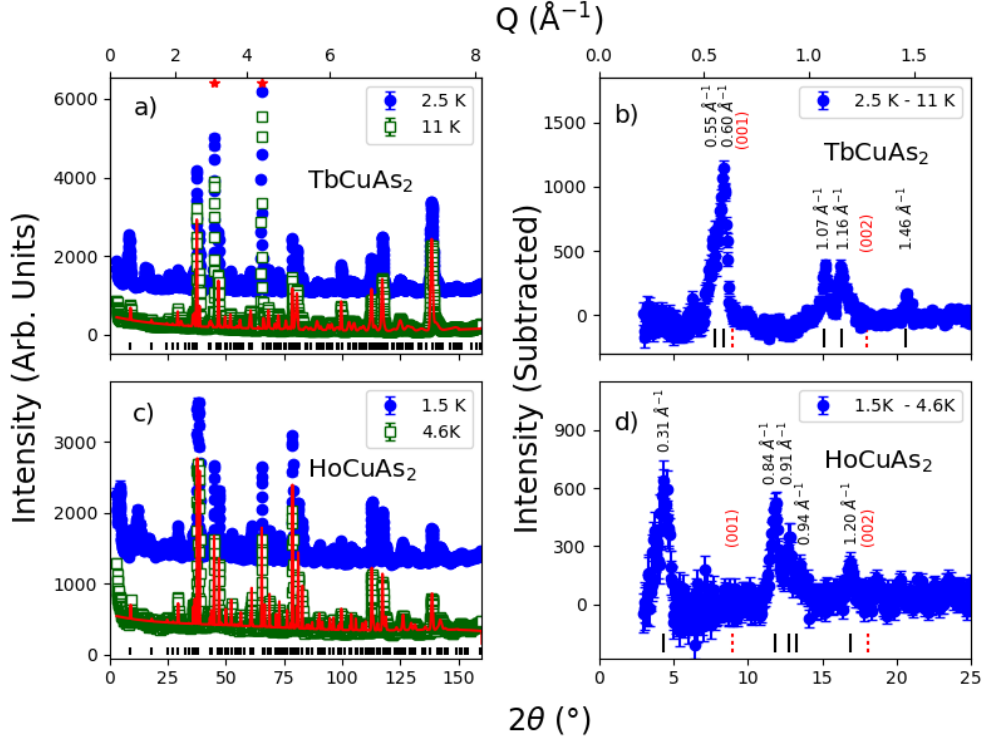

Figure 5 (colour online) (a) High resolution neutron powder diffraction pattern taken on BT1 of TbCuAs$_2$ measured at T = 2.5 K (blue closed circles, vertically offset by 1000) and 11 K (open green squares). (b) Subtracted (magnetic) diffraction pattern of TbCuAs$_2$. (c) High resolution neutron powder diffraction pattern of HoCuAs$_2$ measured at T = 1.5 K (blue closed circles, vertically offset by 1000) and 4.6 K (open green squares). (d) Subtracted (magnetic) diffraction pattern of HoCuAs$_2$. The solid (red) lines in (a) and (c) are the Rietveld structure refinement results for TbCuAs$_2$ and HoCuAs$_2$, respectively. The (red) "*" marks the Al powder peaks from the sample holder with 2θ ≈ 44.9° and 65.3°

### 3.3 Incommensurate Magnetic Order in TbCuAs$_2$ and HoCuAs$_2$

Figure 5 shows the structure and magnetic diffraction pattern of TbCuAs$_2$ and HoCuAs$_2$ measured with the BT1 diffractometer. It is clear that the patterns are considerably more complicated than for the previous materials, with many magnetic Bragg peaks. None of these new Bragg peaks could be indexed with commensurate propagation wave vectors, clearly indicating that the magnetic structures are incommensurate. Using the *K-search* program from *FullProf* suite, we cannot find a unique K-vector for the Tb compound. In particular, the double magnetic peaks near 8° (0.55 and 0.60 Å$^{-1}$) and 16° (1.07 and 1.16 Å$^{-1}$) shown in figure 5(b) preclude any possible solution for a single K-vector. After an extensive search, we found two IC magnetic ordering K-vectors for Tb as $K_{1(Tb)}$ = (0.240, 0.155, 0.48) and $K_{2(Tb)}$ = (0.205, 0.115, 0.28). The magnetic peaks of TbCuAs$_2$ (figure 5(b)) then can be indexed as 0.55 Å$^{-1}$: (000) + $K_1$, 0.60 Å$^{-1}$: (001) – $K_2$, 1.07 Å$^{-1}$: (002) – $K_1$, 1.16 Å$^{-1}$: (002) – $K_2$, 1.46 Å$^{-1}$: (011) – $K_1$. For the Ho system, on the other hand, we obtain the unique result of $K_{(Ho)}$ = (0.121, 0.041, 0.376), together with third order harmonics (i.e. 3*$K$). The HoCuAs$_2$ magnetic peaks (figure 5(d)) then can be indexed as 0.31 Å$^{-1}$: (000)

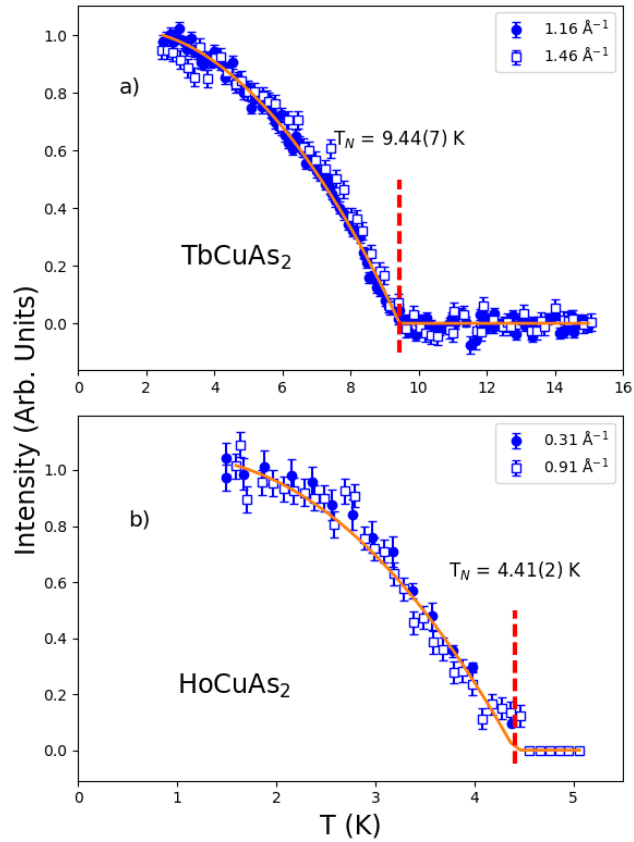

Figure 6 (colour online) Temperature dependence of the integrated magnetic Bragg peak intensities of (a) TbCuAs$_2$ and (b) HoCuAs$_2$, respectively, for two different first-order peaks. The intensities have been scaled to demonstrate that they obey the same temperature dependence. The represented magnetic Bragg peaks occur in (a) TbCuAs$_2$ at Q around 1.16 Å$^{-1}$ (closed circles) and 1.46 Å$^{-1}$ (open squares), and in (b) HoCuAs$_2$ at Q around 0.31 Å$^{-1}$ (closed circles) and 0.91 Å$^{-1}$ (open squares), respectively. The solid (orange) curves show the mean field order parameter fit results for each compound. The vertical dashed (red) lines indicate the fitted $T_N$ = 9.44(7) K for (a) TbCuAs$_2$, and $T_N$ = 4.41(2) K for (b) HoCuAs$_2$, respectively.

+ ***K***, 0.84 Å$^{-1}$: (002) – 3****K***, 0.91 Å$^{-1}$: (001) + ***K***, 0.94 Å$^{-1}$: (000) +3****K***, 1.20 Å$^{-1}$: (102) – 3****K***. The above ***K***-values explain the observed peak positions, with all three components being incommensurate indicating a very complex magnetic structure. For any choices, however, the calculated magnetic scattering intensities cannot reproduce the observed magnetic Bragg peak intensities within the uncertainties with either a simple sine wave, helix or square wave models. Even with arbitrary components of the magnetic moments, we were not able to obtain a completely satisfactory solution for the magnetic structure. Further progress in this work will likely require high quality single crystal measurements.

Figure 6 shows the order parameter measurements for the strongest magnetic peaks of TbCuAs$_2$ and HoCuAs$_2$, measured on BT7. The integrated intensities of these strong fundamental magnetic Bragg peaks evolve smoothly and continuously with temperature, indicating that the magnetic phase transition is 2$^{nd}$ order in nature.

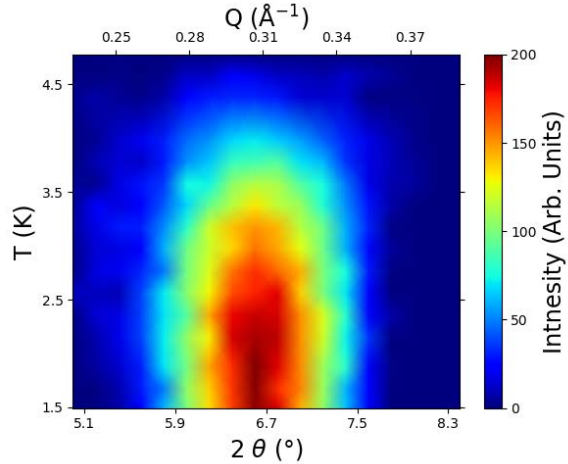

Figure 7 (colour online) Contour map of the magnetic Bragg peak intensity as a function of temperature at a scattering angle around 6.7 degrees for HoCuAs$_2$. For these low-angle measurements we employed an energy analyser and tighter collimation to improve the signal to noise. The background scattering at high temperature has been subtracted. Note that there is no significant change in the wave vector of the incommensurate order, nor is there significant broadening of the scattering in the vicinity of the phase transition.

Mean-field fits of the magnetic Bragg peak intensities give $T_N$ = 9.44(7) K for TbCuAs$_2$ and $T_N$ = 4.41(2) K for HoCuAs$_2$, respectively. Figure 7 shows a colour map of this peak as a function of temperature for HoCuAs$_2$, taken using an energy analyser and Söller collimations of 25´ before and after the sample. We see that the scattering evolves smoothly with temperature without any significant broadening. More importantly, there is no significant variation in the position of the peak, so that the incommensurate wave vector does not vary significantly with temperature.

### 3.4 Absence of Long Range Magnetic Order in YbCuAs$_2$

Figure 8(a) shows the magnetic susceptibility measurements for the YbCuAs$_2$ powder sample, which are similar to those reported previously [10]. This result is indicative of an antiferromagnetic phase transition below $T_N$ = 4 K. However, as seen in figure 8(b), the neutron powder diffraction measurements of the YbCuAs$_2$ sample on BT7 did not show any additional Bragg peaks at T = 1.5 K, well below the presumed $T_N$ = 4 K. Absence of any magnetic Bragg peak indicates either there is no long range magnetic order down to 1.5 K, or the ordered magnetic moment of Yb ions is smaller than our current experimental limit, ≈ 0.2 μ$_B$. It is noteworthy that the authors in Ref. [10] expressed their surprise on the relatively high magnetic ordering temperature of the Yb-compound. Our neutron powder diffraction results could indicate the possibility that the anomalies in resistivity and magnetic susceptibility measurements in YbCuAs$_2$ are related to something else other than the long range magnetic order, as for example, moment instability. There are examples of Yb-compounds where Yb-ions have nearly the theoretical paramagnetic moment but at low temperature the Yb-ions are non-magnetic, for example in, YbMCu$_4$ (M = Ag and In) [20,21].

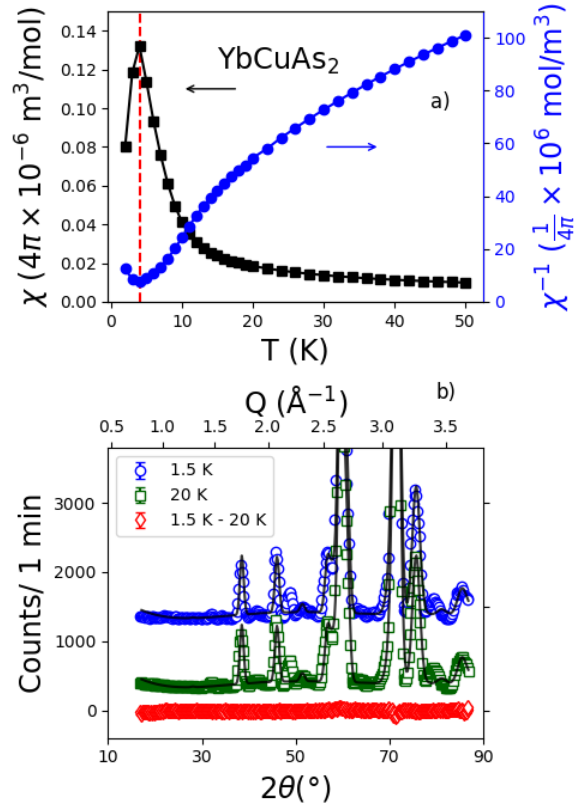

Figure 8 (a) (colour online) Plot of low temperature susceptibility (black squares) for YbCuAs$_2$ and its inverse measured (blue circles) with an applied external magnetic field of $\mu_0 H = 0.5$ T. The peak in the susceptibility at 4 K indicates the magnetic phase transition (T$_N$). (Note that 1 emu/(mol·Oe) = $4\pi \times 10^{-6}$ m$^3$/mol). (b) Diffraction pattern at 1.5 K (blue circles, vertically offset by 1000) and 20 K (green squares). The diamond (red) shows the intensity difference between 1.5 K and 20 K, which does not reveal any magnetic Bragg peaks. The solid (black) lines are the structure refinement results.

We suspect a similar situation may occur in YbCuAs$_2$. Investigations using Mössbauer or μSR techniques might be helpful to clarify this situation, or again if single crystals of sufficient size become available diffraction measurements with more sensitivity could be carried out.

## 4. Crystal Field Effects

The ordered magnetic moments that are well below the free-ion values for some of these materials, the different magnetic moment directions, and incommensurate magnetic structures for others all suggest that crystal field effects are likely important for these rare-earth systems. Consequently, we have carried out inelastic neutron scattering measurements on the Pr, Tb, Ho, and Yb samples to search for crystal electric field (CEF) excitations. The particular aim for these measurements was to determine if there are low-lying CEF states that would be thermally populated near and below T$_N$ and affect the occupancy of the ground state CEF level and thereby the behaviour in ordered state, or in the case of Yb result in a non-magnetic ground state. For these measurements, we employed the BT7 instrument using horizontal focusing of the analyser (since CEF excitations are dispersionless) with a fixed final energy $E_f$ = 14.7 meV, with a PG filter in the scattered beam path to suppress higher order contaminations. The collimation employed was 80´ before (open position) and after the monochromator, an 80´-radial collimator before the analyser, and no collimation after. The measured energy resolution at the elastic position then was 1.4 meV FWHM. Energy transfers were measured from the elastic position up to 30 meV at a number of Q positions within the range from 2.35 Å$^{-1}$ to 3.5 Å$^{-1}$ for selected temperatures above $T_N$. However, we could not identify any CEF excitations in this energy range for any of the measured samples. This suggests that either the observable CEF levels occur above 30 meV, or they are too broad to be observed. There could also be CEF levels at low energies within the instrumental resolution, in the range below ~ 0.5 meV, but it is unlikely that all the CEF levels are in this range for all the materials investigated. These results are in contrast to the previous theoretical predictions of CEF excitations[12], for example the predicted three low-lying state for Tb ions at 2.62, 2.96, and 6.25 meV above the ground state.

## 5. Conclusion

We have carried out systematic neutron diffraction studies on the intermetallic polycrystalline samples of RCuAs$_2$ with R = Pr, Nd, Dy, Tb, Ho, and Yb. We observed long range magnetic order in all the compounds except YbCuAs$_2$, with Néel temperatures of 10 K or below. As inferred from the temperature dependence of the order parameters, the magnetic phase transitions are all continuous (second-order) in nature within the uncertainties of the measurements. We have successfully solved the magnetic structures for the PrCuAs$_2$, NdCuAs$_2$, and DyCuAs$_2$, which all have the same *commensurate* ordering wave-vector K = (0, 0, 0.5). For PrCuAs$_2$, the magnetic moments are parallel to the crystal c-axis, whereas in NdCuAs$_2$ and DyCuAs$_2$ the magnetic moments lie in the a-b plane. In addition, the ordered magnetic moment for PrCuAs$_2$ is much reduced from the free-ion value, while for NdCuAs$_2$ and DyCuAs$_2$, the ordered magnetic moments are close to their free-ions values. These differences in the magnetic moment orientation as well as the value of the ordered moments likely originates from the CEF effects. The TbCuAs$_2$ and HoCuAs$_2$ systems have much more complicated *incommensurate* magnetic structures. The ordering wave-vectors for TbCuAs$_2$ are (0.240,0.155,0.48) and (0.205, 0.115, 0.28), while for HoCuAs$_2$ there is a unique ordering vector of (0.121, 0.041, 0.376) with the 3$^{rd}$ order harmonics. Nevertheless, the magnetic structure refinements using the above IC wave-vectors cannot fully reproduce the intensities of the magnetic Bragg peaks and consequently a complete solution to the magnetic structures has not been obtained. Considering the complexity of the neutron powder diffraction data, measurements on high quality single crystals become essential for the determination of possible magnetic models for TbCuAs$_2$ and HoCuAs$_2$. The absence of any clear magnetic Bragg peaks in YbCuAs$_2$ implies no long range magnetic ordering down to 1.5 K, or an ordered below the limit of 0.2 $\mu_B$.

One possibility is that the Yb CEF ground state is non-magnetic. We were not able to detect any crystal field levels for any of these materials in the energy range from ~0.5 meV to 30 meV, indicating that the levels are too broad to observe or lie at higher/lower energies. Further neutron studies on single crystals of these materials would be helpful to clarify certain details of the magnetic structures and the associated spin dynamics.

**Acknowledgements**

AKG thanks DST (Govt. of India) for funding. GST and ZH thank CSIR and UGC respectively for fellowships. The authors at IIT Delhi thank DST for providing the SQUID facility. The author Y.Z. would like to thank Q. Huang for fruitful discussions. The identification of any commercial product or trade names does not imply endorsement or recommendation by the National Institute of Standards and Technology.